\documentclass[twocolumn,english,aps,reprint, superscriptaddress,showpacs,showkeys]{revtex4-2}
\usepackage{amsmath,amssymb,bbm,mathrsfs,bm,braket,color,graphicx,comment}
\usepackage[colorlinks,citecolor=blue,urlcolor=blue,linkcolor = blue]{hyperref}
\usepackage{soul}
\usepackage{comment}
\usepackage{orcidlink}
\usepackage{bibunits}
\defaultbibliography{references.bib}

\usepackage[mathscr]{euscript}

\usepackage{hyperref}

\usepackage{tikz}

\usepackage{xcolor}
\usepackage{multirow}
\usepackage[normalem]{ulem}

\begin{document}

\title{Vacuum-induced interference in light scattering by multilevel atomic chains}

\author{Aleksei Konovalov}
\email{Aleksei.Konovalov@uibk.ac.at}
\affiliation{Institute for Theoretical Physics, University of Innsbruck, A-6020 Innsbruck, Austria}
\affiliation{Theoretical Physics,  Saarland University,  D-66123  Saarbr\"ucken,  Germany}

\author{Giovanna Morigi \orcidlink{0000-0002-1946-3684}}
\email{giovanna.morigi@physik.uni-saarland.de}
\affiliation{Theoretical Physics,  Saarland University,  D-66123  Saarbr\"ucken,  Germany}
\affiliation{Center for Quantum Technologies (QuTe), Saarland University, Campus, D-66123 Saarbr\"ucken, Germany}

\author{Nicola Piovella \orcidlink{0000-0003-2403-0776}}
\email{nicola.piovella@mi.infn.it}
\affiliation{Dipartimento di Fisica “Aldo Pontremoli”, Università degli Studi di Milano, Via Celoria 16, I-20133 Milano, Italy}

\begin{abstract}
We investigate cooperative light scattering by an ordered chain of multilevel atoms, which possess two quasi-resonant transitions with parallel dipole moments. Interference between dipole transitions induced by coupling with the vacuum gives rise to so called cross-damping and cross-shifts, that modify the incoherent and coherent dynamics and can be manifest in the spectroscopic properties of the emitted light. We determine the excitation spectrum when the atomic chain is driven by an external laser in the limit in which the dipolar transitions can be described by harmonic oscillators and the scattering is coherent. We show that the interplay of multilevel interference and superradiance can give rise to measurable effects in chains of alkali-metal atoms.
\end{abstract}


\maketitle
\section{Introduction}

Cooperative optical effects are one of the cornerstones of quantum optics \cite{Reitz:2022}. When quantum emitters are at relatively high optical densities, they can behave as a collective entity whose optical properties differ from those of isolated atoms \cite{Hepp:1973,gross1982superradiance,Reitz:2022}. Superradiance \cite{dicke1954coherence,gross1982superradiance}, the enhanced cooperative light emission, is perhaps the most prominent example of such collective phenomena. A further prominent phenomenon are the dipole–dipole interactions mediated by multiple scattering processes \cite{gross1982superradiance,O_Dell_2000,Domokos:2013,
olmos2013long, Schuetz:2014, Douglas_2015}. 

In free space, these phenomena are especially pronounced when the interparticle distance is of the order of, or smaller than, the wavelength of the emitters resonant transition \cite{DeVoe:1993, rui2020subradiant, kramer2016optimized, Masson:2024, facchinetti2016storing, jen2016cooperative, bettles2016cooperative}. Combining high-optical densities with spatially-ordered structures, where the medium can form photonic bandgaps \cite{Deutsch:1995,vanCoevorden:1996,Lambropoulos:2000,Rist:2009},
permits one to study the interplay between cooperative phenomena with the classical interference of the Bragg grating \cite{Fernandez-Vidal:2007,Habibian:2014, Reimann:2015,Neuzner_2016}. 

Experimental platforms such as optical lattices \cite{Bloch:2008}, arrays of optical tweezers \cite{Manetsch:2025, barredo2016atom, endres2016atom,huang2023metasurface}, ensembles of emitters in optical resonators \cite{Zeiher:2017,periwal2021programmable, Baum:2022}, enable the realization of optically dense media with spatial periodicity of the order of the wavelength of the resonant light. The spatial long-range order in these systems provides unique conditions for observing quantum interference of scattered light \cite{Fernandez-Vidal:2007,Habibian:2014,jenkins2017many,asenjo2017exponential,shahmoon2017cooperative,bettles2016enhanced,peter2024chirality1, peter2024chirality2,jen2016cooperative,needham2019subradiance,jenkins2012controlled, cech2023dispersionless}. In particular, when one of the lattice periods is a multiple of the transition wavelength, interference effects — and at high-optical density also dipole–dipole interactions — become especially pronounced \cite{rui2020subradiant, kramer2016optimized, Masson:2024, facchinetti2016storing, jen2016cooperative, bettles2016cooperative}. This feature makes such systems promising platforms for exploring phenomena that require strong collective interactions \cite{Mekhov:2009,Rist:2010,kramer2016optimized, perczel2017topological,ferioli2021storage}.

Most theoretical analyses model the emitters with two-level transitions, often including Zeeman sublevels \cite{zhu2016light,bromley2016collective,jennewein2016coherent, peter2024chirality1, peter2024chirality2}, while the role of the multilevel structures, including higher manifold, is relatively unexplored. This is typically justified since the atomic ensembles in most experiments are alkali-metal, such that a closed two-level transition can be realised by optical pumping and sufficiently weak laser excitation. Nevertheless, even in this regime high-precision spectroscopy experiments with single atoms reported features in the excitation spectrum due to off-resonant levels, arising from vacuum-induced interference effects \cite{yost2014quantum,udem2019quantum,horbatsch2010shifts,horbatsch2011shifts}. Vacuum-induced interference refers to interference phenomena between electronic transitions coupled to common modes of the electromagnetic field \cite{ficek2005quantum,Kiffner:2010}. Dipolar electronic transitions satisfying this condition are denoted by parallel dipoles. This interference also occurs when the electromagnetic field modes are in the vacuum \cite{Milonni:1976,Cardimona:1982,Cardimona:1983}, and is thus qualitatively different from interference phenomena of classical emitters. In what follows we will refer to the terms describing this quantum interference as "cross-interference" and, when referring to dissipation, as "cross-damping" terms \cite{yost2014quantum,buchheit2016master}.

In this work we analyse the scattering properties of a chain of multilevel emitters, setting our focus on the interplay between the multilevel interference and Bragg scattering. We analyse the problem in the regime of weak saturation, where scattering is prevailingly coherent and the emitters can be modeled by oscillators \cite{zhu2016light}. By means of an analytical model we highlight the role of vacuum-induced interference in determining the scattering rates. We further analyse numerically the excitation spectra as a function of the chain periodicity for the parameters of the D2 line of $^{23}$Na and of $^{7}$Li atoms. We finally discuss the robustness of vacuum-induced interference against atomic vibrations at the lattice sites, identifying the regime where the effects of vacuum-induced interference can be measurable.

This paper is organized as follows. In Sec.~\ref{section_II}, we introduce the model of coherent dipoles (MCD), formulated for a one-dimensional chain of multilevel quantum emitters. In Sec.~\ref{section_III}, we present an approximate analytic solution of the MCD for a general three-level system within a mean-field approach. Section~\ref{section_IV} is devoted to the excitation spectra of a transversely driven  chain for the parameters of $^{23}$Na or $^{7}$Li atoms. The MCD is solved numerically, and the results are compared with the analytic predictions of Sec.~\ref{section_III}. Finally, in Sec.~\ref{section_V}, we examine the robustness of cross interference effects against atomic vibrations. The conclusions and outlooks are reported in Sec.\ \ref{Sec:Conclusions}. The appendices provide details of the model in Sec.\ \ref{section_II} and of the analytical calculations in Secs. \ref{section_III} and \ref{section_V}.

\section{Theoretical model} \label{section_II}

\begin{figure}[!ht]
	\centering \includegraphics[width=0.6\linewidth]{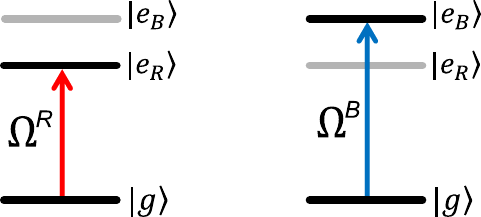}
	\caption{\label{Transition_scheme} Relevant level scheme of each emitter. The colored arrows denote only dipolar transitions which are involved in the excitation process, here labeled by the Rabi frequency multiplied by the corresponding Clebsch-Gordon coefficient. Both dipolar transition are parallel and thus couple with the same modes of the electromagnetic field. Hence, spontaneous decay processes from the excited states can interfere. In this work we assess the magnitude of this interference on coherent scattering when the emitters form a chain. }
    \end{figure}

We consider $N$ identical emitters forming a regular, one-dimensional array. The emitters are localised at the positions $\vec{R}_{\alpha}=\alpha a\,\vec e_x$ (with $\alpha=1,2,...,N$ and $\vec e_x$ the unit vector along $x$) and continuously driven by a monochromatic laser with electric field $\vec{E}_L$ and frequency $\omega_L$. The relevant transitions of each emitter consist of a unique ground state and two excited states at frequency $\omega^i$  ($i=R,B$), the two corresponding dipolar moments $\vec{d}^{\ i}$ are assumed to be parallel. Therefore, the laser drives simultaneously both transitions with Rabi frequencies $\Omega^i=-\vec{d}^{\ i}\cdot\vec{E}_L/(2\hbar)$, as illustrated in Fig.\ \ref{Transition_scheme}. Moreover, photon emission by both transitions can interfere, giving rise to cross-damping and cross-interference dynamics. The interference takes place at the level of individual emitters and also between different emitters, the master equation has been derived in Ref.\ \cite{konovalov2020master} and is reported in Appendix \ref{app_mcd}. 

In this work we assume the weak saturation regime, which allows us to set the ground state population equal to unity and to approximate the dipolar transitions with harmonic oscillators, thus reducing the dynamics to a so-called model of coupled coherent dipoles \cite{zhu2016light,bromley2016collective, jennewein2016coherent,jennewein2018coherent}. In this limit, we denote by $b_{\alpha}^i$ the optical coherence of the dipolar transition $i$ of the emitter at position $\vec{R}_\alpha$, while all other emitters are in the ground state. The equations of motion take the form 
\begin{eqnarray}  
		\dot b_{\alpha}^i 
		&=& 
		\sum_{j=R,B}b_{\alpha}^j
        \left(
        \mathrm{i} \Delta^{ij}  
        -
		\frac{\Gamma^{ij}}{2}		 
        \right)
        -
		\mathrm{i} \Omega_{\alpha}^{i}-
		\sum_{\beta\ne \alpha\,,1}^{N}  \sum_{j}G_{\alpha\beta}^{ij}  
		b_{\beta}^j\,,\nonumber\\\label{eq:b}
\end{eqnarray}
see Appendix \ref{app_mcd} for details. Here, the first two terms on the right-hand side (RHS) are the single-emitter contributions, while the last term on the RHS accounts for collective effects. We first discuss the single-emitter part. Here, for $i=j$, then $\Gamma^{ii}\equiv\Gamma^i$  is the linewidth of the dipolar transition $i$, $\Delta^{ii}\equiv\Delta^{i}_L=\omega_L-\omega^i$ is the laser detuning from the dipolar transition, and $\Omega^i_{\alpha}\equiv\Omega^i{\rm e}^{{\rm i}\vec{k}_L\cdot \vec{R}_\alpha}$ is the Rabi frequency for the dipolar transition $i$, which depends on the position $\vec{R}_\alpha$ through the phase of the running wave ${\rm e}^{{\rm i}\vec{k}_L\cdot \vec{R}_\alpha}$. The terms $\Gamma^{RB},\Gamma^{BR}$ and $\Delta^{RB},\Delta^{BR}$ give rise to cross-damping and interference, respectively, induced by the coupling with vacuum and thermal fluctuations of the electromagnetic field modes \cite{ficek2005quantum,buchheit2016master}. Note that we have discarded the center-of-mass motion. Moreover, we assume optical transitions and neglect thermal effects, since the mean photon number at optical frequencies and room temperature is negligible \cite{konovalov2020master}. 

The last term on the RHS describes the interatomic coupling via the coherent and incoherent part of the dipolar interaction through the Green's function:
\begin{multline} \label{eq:GF}
G_{\alpha\beta}^{ij} =  \frac{3\tilde\gamma^{ij}_{\alpha\beta}}{4} \frac{e^{\mathrm{i}kR_{\alpha\beta}}}{\mathrm{i}kR_{\alpha\beta}}
\left\{
 \left[ (\vec{e}_{\alpha}^{\,i*} \cdot \vec{e}_{\beta}^j ) - (\vec{e}^{\,i*}_\alpha\cdot\vec{e}_x) (\vec{e}^{\ j}_\beta\cdot\vec{e}_x)\right]  \right.
\\
\left.
+
\left( \frac{1}{kR_{\alpha\beta}} - \frac{1}{\mathrm{i}(kR_{\alpha\beta})^2} \right)
\left[ (\vec{e}_{\alpha}^{\,i*} \cdot \vec{e}_{\beta}^{\ j} ) - 3(\vec{e}^{\,i*}_\alpha\cdot\vec{e}_x)(\vec{e}^{\, j}_\beta\cdot\vec{e}_x) \right]
\right\}
\end{multline}
where $R_{\alpha\beta} = |\alpha-\beta|a$ 
is the linear distance between two emitters located at $\alpha a$ and $\beta a$, while $\vec{e}_{\alpha}^{\ i}=\vec{d}^{\,i}_{\alpha}/d^i_{\alpha}$ is the unit vector along the direction of the corresponding dipolar transition and $\vec{e}_x=\vec{R}_\alpha/|\vec{R}_\alpha|$. Note that since the dipolar transitions are parallel, then $\vec{e}_\alpha^{\ i*}\cdot \vec{e}_\beta^{\ j}=1$. 
The parameter scaling Eq.\ \eqref{eq:GF} has the dimension of a rate and reads:
$$ \tilde\gamma^{ij}_{\alpha\beta}= \frac{4}{3} \frac{d^{i*}_{\alpha} d^{j}_{\beta}}{\hbar c^3} \left(\frac{\omega^i+\omega^j}{2}\right)^3\mathcal{F}^{ij}\,.$$
It describes the strength of dipolar transitions of different atoms, including the effect of identical transitions ($i=j)$ as well as different transitions ($i\neq j$), and is scaled by the function $\mathcal{F}^{ij}$, which results from the systematic derivation of the coarse-grained master equation, see Appendix \ref{app_mcd}. For the parameters considered in this work, $\mathcal{F}^{ij}\approx 1$. Thus, for $i=j$ then $\tilde\gamma^{ii}_{\alpha\beta}=4 |d^{i}|^2(\omega^i/c)^3/3\hbar$ takes the form well known for two-level emitters \cite{gross1982superradiance}. For $i\neq j$, the rate $\tilde\gamma^{ij}_{\alpha\beta}$ depends on the average value between the two transition frequencies. We remark that the validity of the coherent-dipoles model in Eq.\ \eqref{eq:b} is restricted to the weak saturation regime, for $N\Omega^i\ll |{\rm i}\Delta^i_L-\Gamma^i/2|$.

In what follows, we will focus on chains composed by alkali-metal atoms. In this case, the assumption of a three-level transition, as we postulated in this section, is an oversimplification that, amongst others, artificially breaks the spherical symmetry of each atom. In fact, this symmetry leads to vanishing cross-interference and cross-damping terms of the individual atoms, that is, $\Gamma^{i\neq j}=0$ and $\Delta^{i\neq j}=0$. On the other hand, the spatial order leads to a finite contribution of the cross-interference and damping terms between different atoms. In order to single out their dependence on the spatial periodicity and determine their order of magnitude, in the rest of this paper, we keep the three level configuration but set $\Gamma^{i\neq j}=0$ and $\Delta^{i\neq j}=0$ in Eq.\ \eqref{eq:b}. We refer the reader to Ref.\ \cite{konovalov2020master} for a full numerical analysis of the interplay of single- and multi-emitter cross-interference in the excitation spectra of two emitters.

Our goal is to identify the role of cross-interference in the photon-count signal (also known as the excitation spectrum) integrated over the full detection angle. In the low saturation regime, it takes the form:
\begin{multline} \label{PCS_general}
S_{\Omega}(\delta_L) =\Gamma\sum_{\alpha=1}^N\sum_{i=R,B} |b^i_\alpha|^2
\\
+\sum_{\alpha=1}^N\sum_{\beta=1\atop \beta\neq\alpha}^N\sum_{i,j=R,B} 2\text{Re}\left[ G_{\alpha\beta}^{ij}\right]\, b_{\alpha}^i  b_{\beta}^{j\,*}, 
\end{multline}
where $b_{\alpha}^i$ is calculated in steady state, after setting $\dot{b}_\alpha^i=0$ in Eq.\ \eqref{eq:b}.

\section{Mean-field solution} \label{section_III}

We first consider the stationary solutions of Eqs. \eqref{eq:b} in the mean field limit, which we specify below. This will provide insight into how cross-interference between non-degenerate, yet parallel, dipoles influences inter-atomic interactions. 

\subsubsection{Coherences}
For simplicity, we assume $\Gamma_R=\Gamma_B=\Gamma$. We set $\dot b_i^\alpha=0$ and introduce the variable $\tilde b_\alpha^i=b_\alpha^i\exp(\mathrm{i}\vec{k}_L\cdot\vec{R}_{\alpha})$, which satisfies the coupled equation
\begin{equation} \label{eq:MCD_chain}
\tilde b_{\alpha}^i
=
\frac{ \Omega^{i} }{\mathrm{i}\mathcal{D}^i} 
-
\frac{1}{\mathcal{D}^i}
\sum_{\beta(\ne\alpha)}^{N} \sum_{j=R,B}e^{-\mathrm{i}\vec{k}_L\cdot\vec{R}_{\alpha\beta}}  
G_{\alpha\beta}^{ij}\,\tilde b_{\beta}^j, 
\end{equation}
where $\mathcal{D}^i=\Gamma/2 - \mathrm{i}\Delta^i$. For sufficiently dilute atomic ensembles, we assume that each atom is equally affected by the entire ensemble and  neglect edge effects, which is a valid approximation for sufficiently long arrays. In this regime, we denote the coherence $\tilde b_\alpha^i=b^i$, such that 
\begin{equation} \label{b(n)mf}
b^i
=
\frac{ \Omega^{i} }{\mathrm{i}\mathcal{D}^i} 
-
\frac{1}{\mathcal{D}^i}
\sum_{j=R,B}\left\langle G^{ij} \right\rangle b^j, 
\end{equation}
where the mean-field Green function reads
\begin{equation} \label{mf_condition}
\left\langle G^{ij} \right\rangle=
\frac{1}{N} \sum_{\alpha}^{N} \sum_{\beta(\ne \alpha)}^{N} 
e^{-\mathrm{i} \vec{k}_L\cdot\vec{R}_{\alpha\beta}} G_{\alpha\beta}^{ij}\,.
\end{equation} 

\noindent The formal solution of \eqref{b(n)mf} takes the form:
\begin{subequations} \label{coh_solution} 
	\begin{align}   \label{coh_solutionR}
b^R &= - \mathrm{i}\frac{\Omega^R(\mathcal{D}^B+\langle G^{BB}\rangle)-\langle G^{RB}\rangle \Omega^B}
{(\mathcal{D}^R + \langle G^{RR})\rangle)(\mathcal{D}^B +\langle G^{BB}\rangle)- \langle G^{RB} \rangle\langle G^{BR}\rangle},
\\
b^B &= - \mathrm{i}\frac{\Omega^B(\mathcal{D}^R+\langle G^{RR}\rangle)-\langle G^{BR}\rangle \Omega^R}
{(\mathcal{D}^B + \langle G^{BB})\rangle)(\mathcal{D}^R +\langle G^{RR}\rangle)- \langle G^{RB} \rangle\langle G^{BR}\rangle}, \label{coh_solutionB}
\end{align}
\end{subequations}
where each mean-field coherence contains variables associated with the other transition as well. 
If we formally set the cross-interference terms to zero, $\langle G^{RB} \rangle =0$ and $\langle G^{BR} \rangle =0$, then both coherences recover the known result for systems of emitters with no cross-interference terms between dipoles, $b^i=\frac{-\mathrm{i} \Omega^i}{\mathcal{D}^i + \langle G^{ii} \rangle}$ \cite{zhu2016light}. 

The solutions \eqref{coh_solution} are proper rational functions of the laser detuning and can be written as a sum of simple fractions, whose denominators determine the spectrum of the system.  In further analysis of the solutions \eqref{coh_solution} we will see that the result depends on whether the driving laser is tuned close to either the lower- or higher-frequency transition. Hereafter, we refer to the transition, at which the laser is tuned close to resonance, as {\it driven}, while the other, distant by the frequency gap $\delta \omega$ (with $|\delta_L|\ll\delta\omega$) we call {\it interfering}.   

We first investigate the case where the lower-frequency $R$-transition is driven, that is, the laser frequency $\omega_L = \omega^R + \delta_L$ with $|\delta_L| \ll \delta\omega$. Similarly, the $B$-transition is the interfering one, so that $\mathcal{D}^R = \Gamma/2 - \mathrm{i}\delta_L$ and $\mathcal{D}^B = \Gamma/2 + \mathrm{i}(\delta\omega - \delta_L)$, where $\delta\omega=\omega_B-\omega_R$.
Then, up to the first order correction in $\epsilon=\Gamma/\delta\omega\ll 1$,
\begin{subequations} \label{MFcoh_A}
\begin{align} 
b^R_{\rm driven} 
&= b^{(0)R}+\epsilon b^{(1)R}\\
b^B_{\rm interfering}&=\epsilon b^{(1)B}
\end{align}
\end{subequations}
with the leading order term 
\begin{equation}
b^{(0)R}= -\mathrm{i}\frac{\Omega^R}{\Gamma^R/2-{\rm i}\delta_L}
\end{equation}
describing the coherence for a collective excitation, with damping rate ($j=R,B$)
\begin{equation}
\label{Gamma:SR}
\Gamma^j=\Gamma+2\langle G^{jj}\rangle\,.
\end{equation}
Hence, at lowest order the emitters behave as two level transitions.
The terms in first order read
\begin{eqnarray} 
&&b^{(1)R}=\frac{\langle G^{RB} \rangle}{\Gamma^R/2-{\rm i}\delta_L}\left(\frac{\Omega^B}{\Gamma}- \frac{\Omega^R\langle G^{BR}\rangle}{\Gamma(\Gamma^R/2-{\rm i}\delta_L)}\right)\\
&&b^{(1)B}=-\frac{1}{\Gamma}\left(\Omega^B-\langle G^{BR}\rangle\frac{\Omega^R}{\Gamma^R/2-{\rm i}\delta_L} \right)\,,
\end{eqnarray}
and include off-resonant excitation of single-atom transitions (first term on the RHS of $b^{(1)B}$) and multilevel interference (proportional to $G^{RB}$ and/or 
$G^{BR}$
).

Similarly, when the laser is tuned near the higher-frequency $B$-transition, $\omega_L = \omega^B + \delta_L$, we obtain the solution by swapping the index $R\leftrightarrow B$ and the sign $\epsilon\to-\epsilon$ in Eqs.\ \eqref{MFcoh_A}.

\subsubsection{Cross interference effects on the linewidth and frequency shift}
\label{Sec:CLS-CDR}
Let us now discuss the spectral properties of the coherences \eqref{coh_solution}, they can be extracted after inspecting the denominators. At first order in $\epsilon$ the linewidth is given by
\begin{equation}
\langle \Gamma_{\text{CDR}}^R \rangle \label{CDR_R}
=
\Gamma +  2\text{Re} \left[ \langle G^{RR} \rangle  + \mathrm{i} \frac{\langle G^{RB} \rangle \langle G^{BR} \rangle}{\delta \omega} \right],
\end{equation}
and contains the superradiant contribution of two-level emitters known as Cooperative decay rate (CDR) as well as the cross-interference terms. The analogous expression for the lineshift reads:
\begin{equation}
\langle \Lambda_{\text{CLS}}^R \rangle \label{CLS_R}
=
\text{Im} \left[ \langle G^{RR} \rangle  + \mathrm{i} \frac{\langle G^{RB} \rangle \langle G^{BR} \rangle}{\delta \omega} \right],
\end{equation}
where the first term is the linear contribution to the so-called Collective Lamb Shift (CLS) \cite{putnam2016collective, jennewein2016coherent, rohlsberger2010collective, bromley2016collective,glicenstein2020collective}, while the second term is the contribution due to cross-interference. Note that these cross-interference contributions have opposite influence on the lower- and higher-frequency transitions due to the sign inversion of the energy gap parameter $\delta \omega$ in Eqs.\ \eqref{CDR_R}-\eqref{CLS_R}.

\subsubsection{Excitation spectrum}

We now find an analytical form of the photon count signal for the mean field solution. From Eq.\eqref{PCS_general} we obtain:
\begin{equation}
\label{S:omega_MCD}
S(\delta_L)=\Gamma\sum_{i=B,R}|b^i|^2 + \sum_{i,j=B,R}2{\rm Re}\big[\langle G^{ij} \rangle\big]\, b^{i}b^{j*},
\end{equation} 
where  $b^{i}$ are given by Eqs.\ \eqref{coh_solution}.
Using Eq. \eqref{MFcoh_A}, the excitation spectrum, up to the first order in $\epsilon$, is the sum of two terms ($j=R$ or $B$): 
\begin{equation}
\label{S:approx}
S_j(\delta_L) =S_j^{(0)}(\delta_L)+\epsilon S_j^{(1)}(\delta_L)\,,
\end{equation}
where the zero order term, for $j=R$,
\begin{equation}
\label{S:0}
S_R^{(0)}(\delta_L)=\big(\Gamma + 2 {\rm Re} \big[ \langle G^{RR} \rangle \big] \Big)|b^{(0)R}|^2\,,
\end{equation}
is the excitation spectrum in the absence of cross interference. Equation \eqref{S:0} has a Lorentzian shape centered on $\omega_R+{\rm Im}\{\Gamma_R\}$ with linewidth ${\rm Re}\{\Gamma_R\}$. The contribution of first order in $\epsilon$ modifies the Lorentzian profile and has the form
\begin{align}
S_R^{(1)}(\delta_L)=2\Big( \Gamma + & 2 {\rm Re} \big[ \langle G^{RR} \rangle\big]{\rm Re}\big[ b^{(0)R}b^{(1)R*}\big]\big) \nonumber
\\
+&
4{\rm Re}\big[\langle G^{RB} \rangle\big]{\rm Re}\big[ b^{(0)R}b^{(1)B*}\big].
\end{align}
The excitation spectrum $S_B(\delta_L)$ close to the resonance at frequency $\omega_B$ is found analogously.

\section{Excitation spectrum of the chain} \label{section_IV}

We now evaluate the excitation spectra for a periodic chain of atoms aligned along $x$ and with period $a$. The atoms are continuously driven by a  $\pi$-polarized laser whose polarization and wave vector are both oriented perpendicular to the chain axis. The parameters we consider are the ones characterising the two parallel transitions of the D2 line of (i) sodium $^{23}$Na and (ii) of lithium $^{7}$Li, namely, the transitions between the ground state $S_{1/2}$ with ${F=2,\,M_F=2}$ and the excited states $P_{3/2}$ with quantum numbers ${F'=2,\,M_{F'}=2}$ and $ {F'=3,\,M_{F'}=2}$, as detailed in App.\ \ref{app:atomic_levels}. We take two types of atoms to compare the influence of the cross-interference terms, which, as we can see from the results of the previous section, depend on the magnitude of the ratio $\epsilon=\Gamma/\delta \omega$. We note that $\epsilon = 0.17$ for sodium  \cite{steck2003sodium} while $\epsilon = 0.65$ for lithium \cite{zelener2015magneto}. 
\begin{figure}[!ht]
	\centering \includegraphics[width=0.74\linewidth]{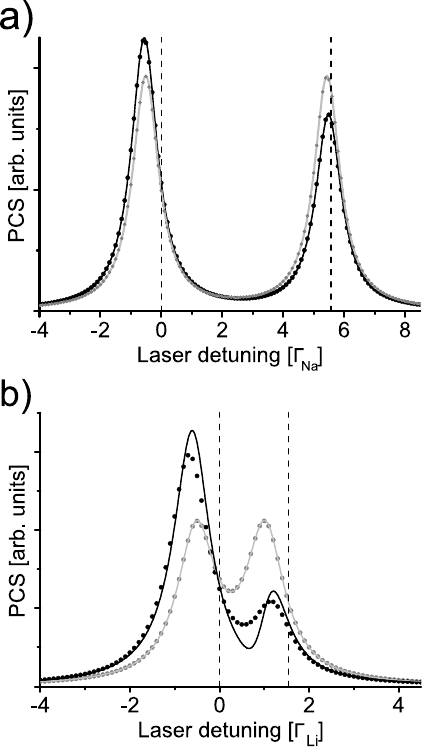}
	\caption{\label{signals} The black dots are the photon-count signal, Eq.\ \eqref{PCS_general}, vs laser detuning for a chain of $N=1000$ emitters consisting of two parallel dipoles with the parameters of (a) $^{23}$Na and (b) $^{7}$Li D2 line. The interparticle distance is $a=\lambda$, with $\lambda$ the wavelength of the optical transition. Vertical dashed lines indicate the bare transition frequencies, the laser detuning is defined with respect to the lower-frequency transition. Solid lines correspond to the result in perturbation theory in $\epsilon$ the grey solid line is found by setting all cross interference terms to zero ($S_R^{(0)}+S_B^{(0)}$, see Eq.\ \eqref{S:0}). The grey dots are the results of a simulation of the chain by setting artificially $G^{RB}=0$ in Eq.\ \eqref{PCS_general}). The black solid line includes the first order corrections ($S_R+S_B$, see Eq.\ \eqref{S:approx}) and qualitatively reproduces the numerical results summing up the individual contribution of the emitters along the chain.
	} 
\end{figure}

The spectrum, Eq.\ \eqref{PCS_general}, is calculated by numerically solving the coupled equations \eqref{eq:MCD_chain} for various values of laser detuning. Figure \ref{signals} displays the result for a chain of 1000 emitters with the parameters of Sodium (a) and Lithium (b) D2 line. Each signal consists of two peaks, shifted from the bare atomic transition frequencies. The result is compared with the mean-field result at zero and first order in $\epsilon$. A net discrepancy between numerical and perturbative results is visible in subplot (b), where $\epsilon=0.65$. This discrepancy can be all attributed to the effect of the cross-interference between multiple dipoles.  In fact, as shown in the previous section,  the smaller the energy gap between interfering dipoles, the stronger the impact of cross-interference. Moreover, cross-interference contributes to the $R$- and $B$-transitions in seemingly opposite ways, which is formally reflected by the sign reversal of the energy gap, $\delta\omega \to -\delta\omega$. Interestingly, the model which includes the cross-interference effects at first order in $\epsilon$ qualitatively reproduces the numerical results. 

We now discuss the properties of the spectra as a function of the chain period. We numerically determine the spectrum on large chains for different values of $a$. We focus on the frequency shift and on the linewidth of the resonances with respect to the ones of the bare atom and we determine their value by fitting each spectrum close to the corresponding resonance. Based on the analytical results, Eq.\, \eqref{S:approx}, the numerical signals are fitted with Fano-like curves
\begin{equation} \label{fano_fit}
S_{\text{fit}}=\sum_{i=R,B} \left(\frac{a_i}{\gamma_i^2/4+(\Lambda_i-\delta_L)^2} + \delta_L\frac{b_i}{\gamma_i^2/4+(\Lambda_i-\delta_L)^2}\right)\,, 
\end{equation}
where the parameters $\{\gamma_i \}_{i=R,B}$ and $\{\Lambda_i \}_{i=R,B}$ correspond to the linewidths and line positions. Below we label the linewidth by CDR and the line shift by CLS.  

\begin{figure}[!ht]
	\centering \includegraphics[width=1.0\linewidth]{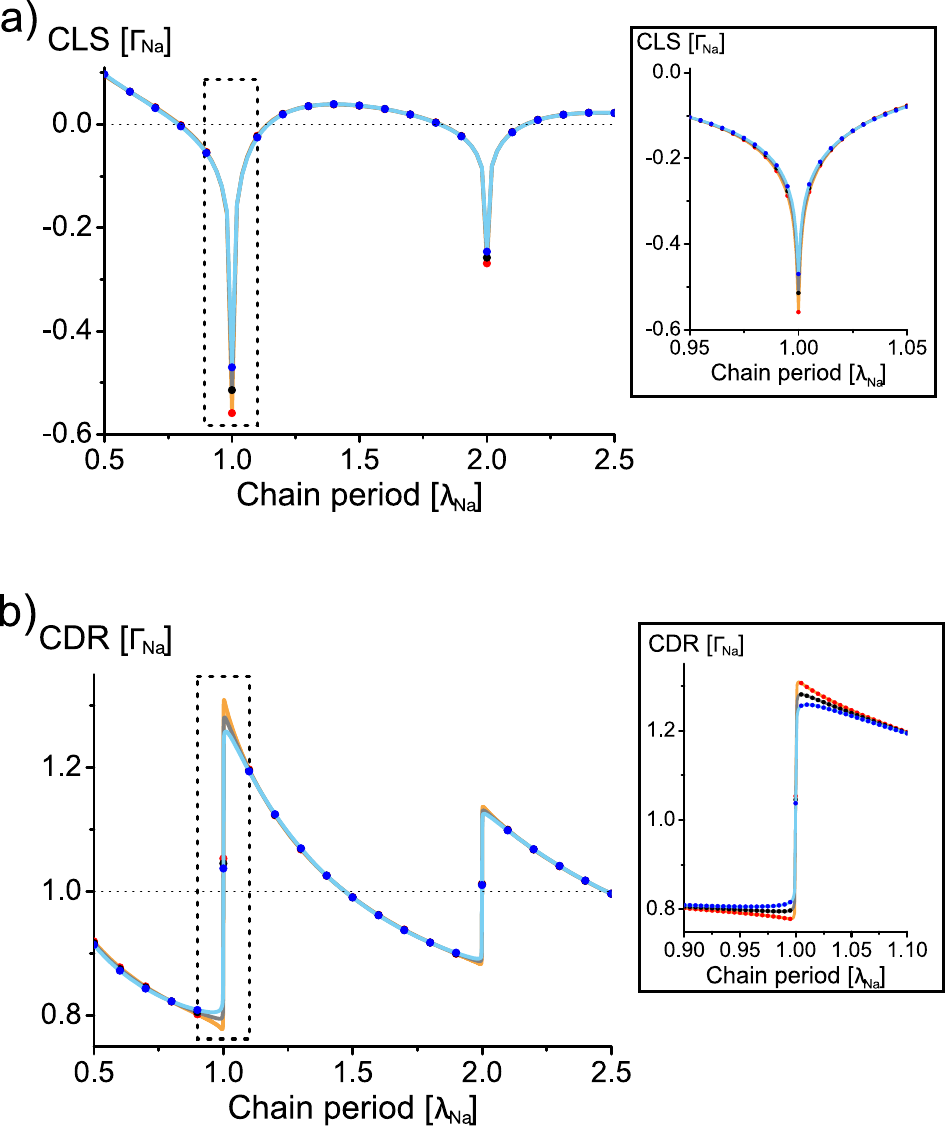}
	\caption{\label{Na_parameters_VS_period_N=1000} Onset: a) Line shift $\Lambda_i$ (CLS) and b) decay rate $\gamma_i$ (CDR) as functions of the chain period and in units of $\Gamma$. The dots are the results of numerical integration of the excitation spectra for a chain of $N=1000$ three-level emitters with the parameters of $^{23}$Na and fitting using \eqref{fano_fit}. The orange and cyan lines are the analytical results for the $R$- and $B$-transitions, respectively, as in Sec.\ \ref{Sec:CLS-CDR}. For reference, the grey line shows the result when the cross-interference terms are set to zero in the model. It significantly overlaps with the predictions of the full model, except at periodicity $a\simeq \lambda, 2\lambda$. The insets display zooms into the region inside the corresponding box of the onset. Here, the predictions of the full model with cross-interference is compared with the one of a model where cross-interference terms are set to zero (black). The red and blue dots refer to the $R$- and $B$-transitions, respectively.
	} 
\end{figure}
\begin{figure}[!ht]
	\centering \includegraphics[width=1.0\linewidth]{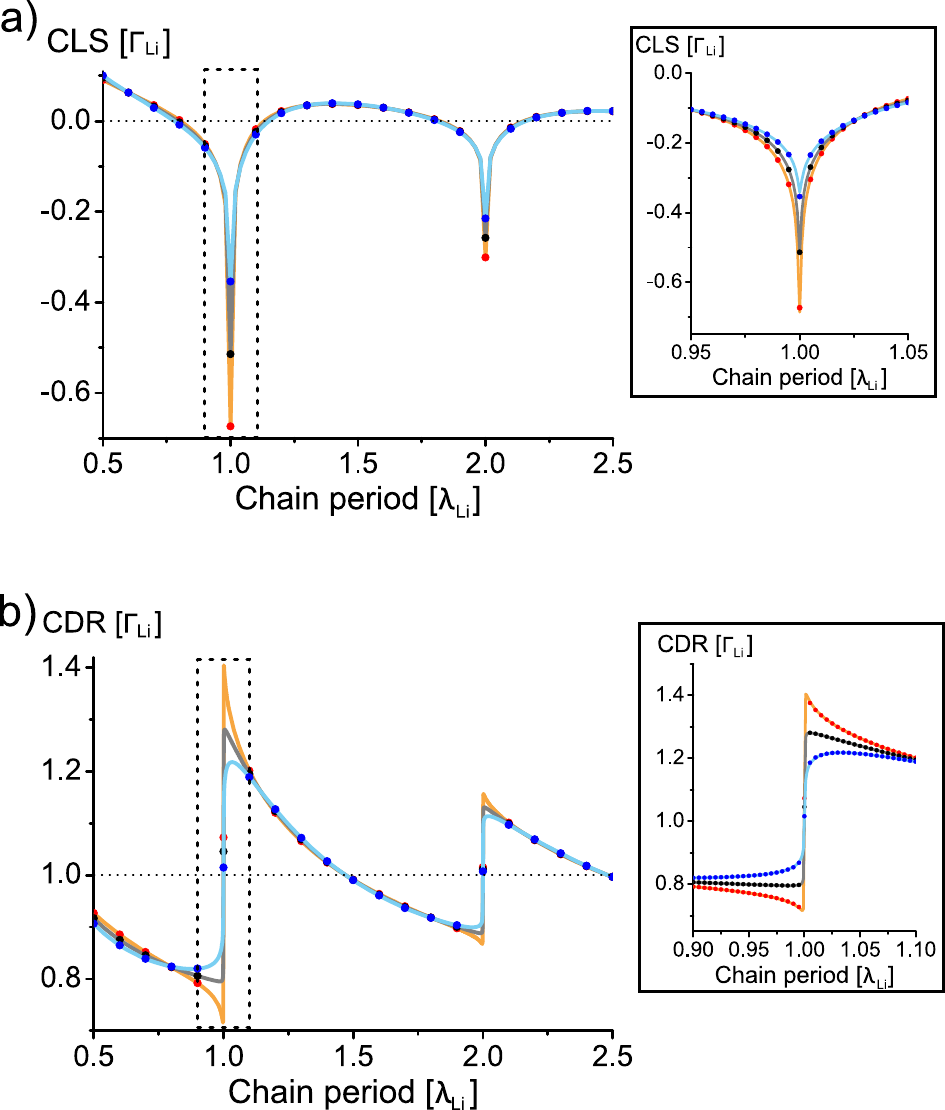}
	\caption{\label{Li_parameters_VS_period_N=1000} Same as Fig.\ref{Na_parameters_VS_period_N=1000} but for the parameters of $^{7}$Li. 	} 
\end{figure}

Both analytic and numerical results are presented in Fig. \ref{Na_parameters_VS_period_N=1000} for the parameters of sodium and in Fig. \ref{Li_parameters_VS_period_N=1000} for the parameters of lithium and for periodicities $a>\lambda/2$.
The behavior of the CDR as a function of the chain period resembles a saw-tooth profile, while the corresponding dependence for the CLS is comb-like. These results are consistent with those reported for chains of two-level emitters, see \cite{bettles2016cooperative, kramer2016optimized, masson2020many, piovella2024cooperative, nienhuis1987spontaneous}. For chains with periods that are integer multiples of the transition wavelength, the effect of cross-interference becomes visible. Moreover, significant differences appear between the predictions of the two- and three-level mean-field models, namely, the models without and with cross-interference terms. The discrepancy is especially pronounced in the depth of the CLS peaks. For lithium, the impact of cross-interference is considerably stronger than for sodium, consistent with the photon-count signal results shown in Fig. \ref{signals}. Notably, for chain period $a=\lambda$, the effect of cross-interference on the CLS peak depth is comparable to that of the CLS itself.

\begin{figure}[!ht]
	\centering \includegraphics[width=1.0\linewidth]{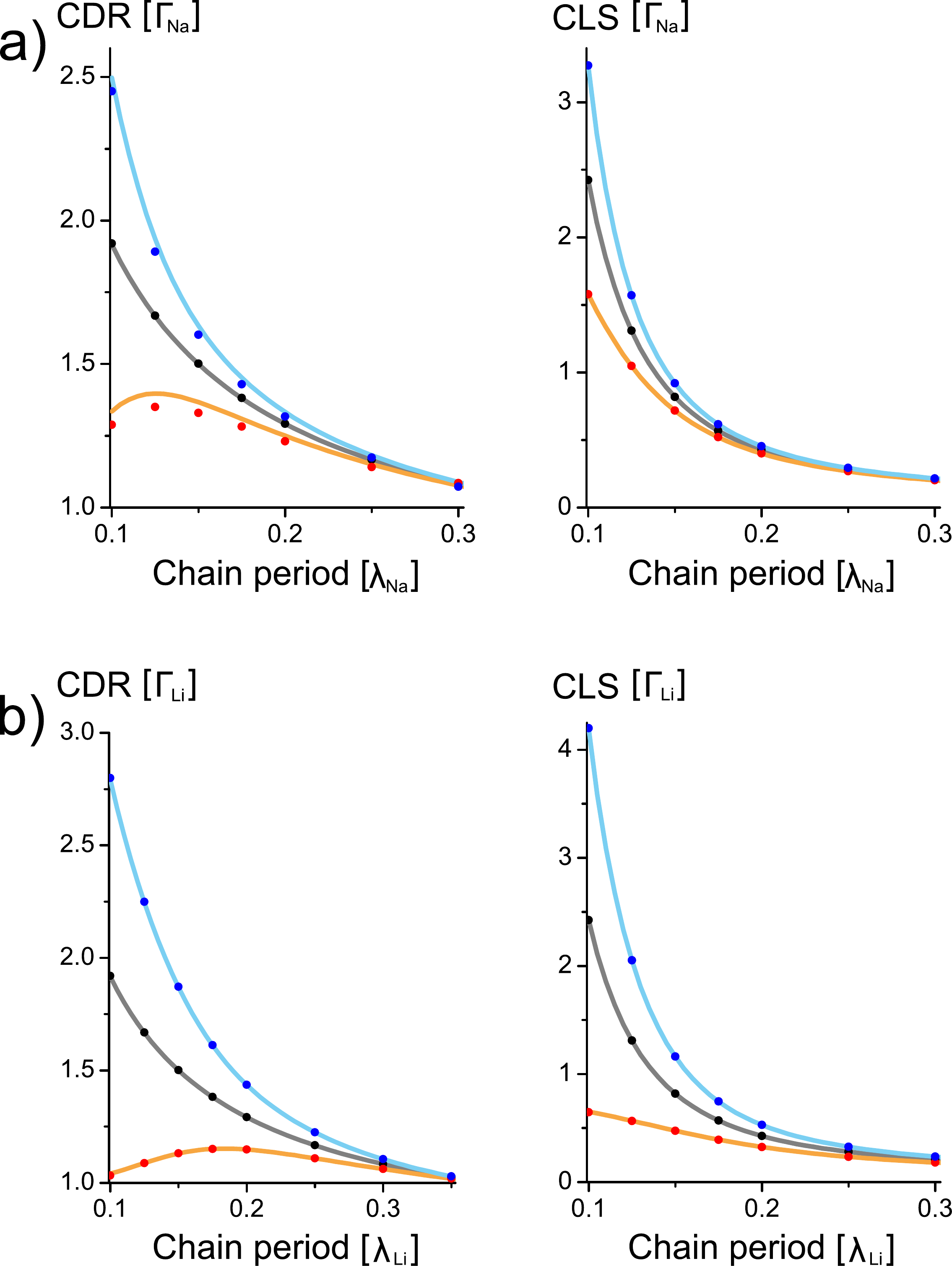}
	\caption{\label{SD} CLR and CLS for the parameters of (a) Sodium and (b) Lithium and $a<\lambda/2$. The parameters and legenda are the same as in Figs.\ \ref{Na_parameters_VS_period_N=1000} and \ref{Li_parameters_VS_period_N=1000}. 
	} 
\end{figure}
Figure \ref{SD} displays both linewidths and decay rates for periodicity in the subwavelength regime. The effects of cross-interference become increasingly pronounced as $a$ decreases. For periods $a \lesssim 0.15 \lambda $, the influence of cross-interference terms leads to significant deviations from the prediction of a model, that discards these effects (grey line). These results highlight the necessity of including these effects in optically dense ensembles. In the following, we discuss the influence of residual atomic motion on the signal.

\section{Impact of atomic motion} \label{section_V}

So far, we have assumed an idealized situation in which the atoms are fixed at given positions. We now include the effect of atomic motion by means of a phenomenological model in order to assess the impact on the excitation spectrum, and in particular, on the determination of the line shift due to cross-interference.
We assume that the atoms are tightly bound at the positions $\vec{R}_{\alpha}$ and perform uncorrelated fluctuations. We denote the displacement from $\vec{R}_{\alpha}$ by $\vec{r}_{\alpha}$. The density distribution of the atom at $\vec{R}_{\alpha}$ is then given by
\begin{equation}
\label{Gauss}
n_\alpha(\vec{r}_{\alpha})
=
\frac{1}{(\sqrt{\pi} \Delta)^3} e^{-\frac{(\vec{r}_{\alpha})^2}{\Delta^2}}\,,
\end{equation}
with $\Delta$ the width of the distribution, $\Delta\ll a,\lambda$. In this regime, $k\Delta\ll 1$. By performing a  convolution of the Green's function \eqref{eq:GF} at the positions of the atoms with their spatial distribution about $\vec{R}_\alpha$, we obtain
\begin{equation} \label{Noisy_GF}
\bar{G}_{\alpha\beta}^{ij}
\approx
\left( 1-\frac{1}{2}(k\Delta)^2 \right) 
G_{\alpha\beta}^{ij}(kR_{\alpha\beta} ),
\end{equation}
where we neglected terms $\sim \frac{(k\Delta)^2}{(kR)^3}, \frac{(k\Delta)^2}{(kR)^4}$, since they are of higher order. This result leads to a rescaling of the CDR and of the CLS by the factor $\left( 1-\frac{1}{2}(k\Delta)^2 \right)$.

\begin{figure}[!ht]
	\centering \includegraphics[width=0.8\linewidth]{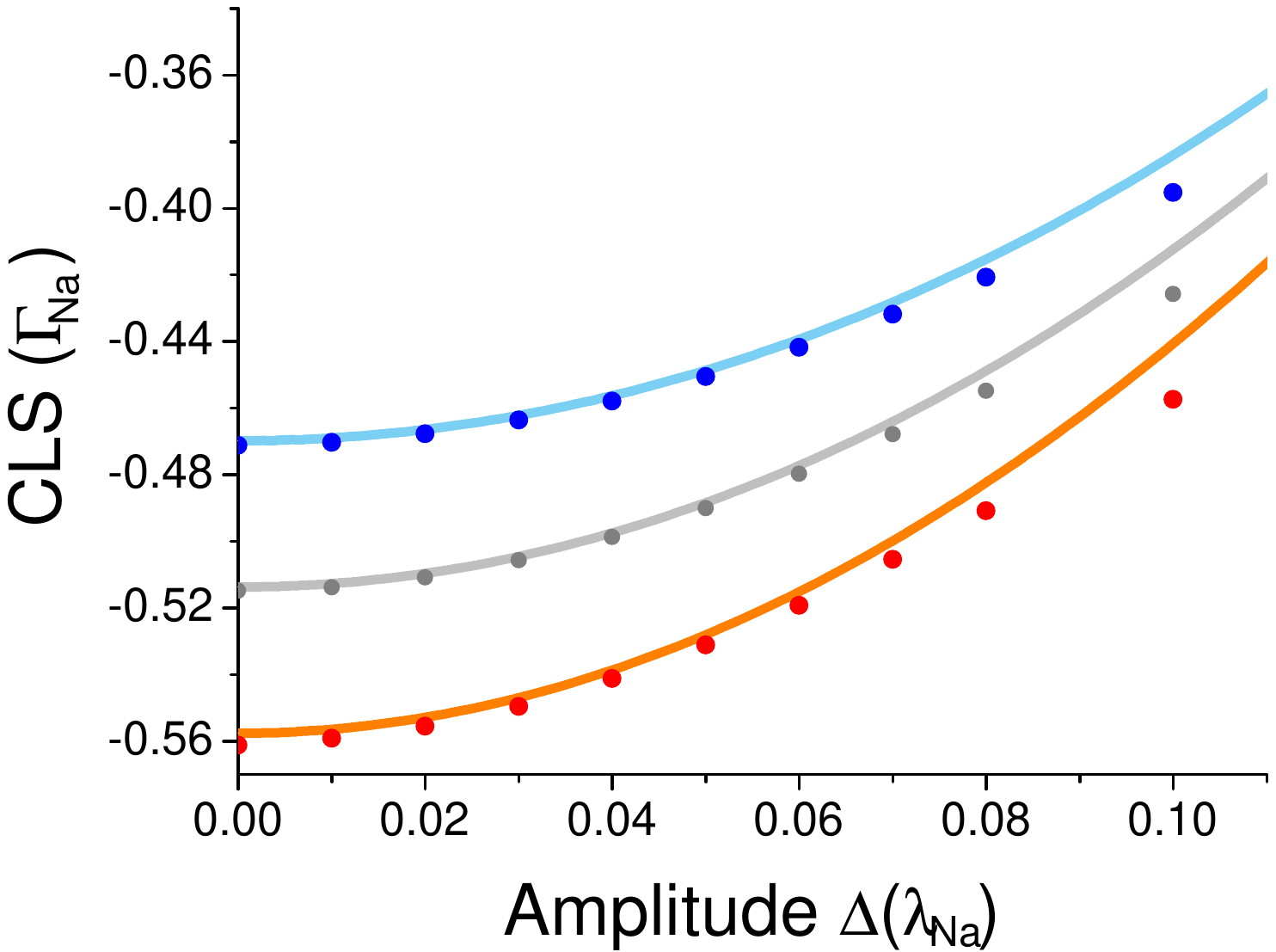}
	\caption{\label{CLS_Na_Delta} The shift CLS for a chain with period $a=\lambda$ versus the position fluctuation amplitude $\Delta$. The dots are obtained by solving numerically Eq.\ \eqref{PCS_general} for a chain of $N=1000$ Sodium atoms. The solid lines are the analytic predictions of the mean-field model, rescaled by the factor $(1-(k\Delta)^2/2)$ (orange for $R$-, cyan for $B$-transition). The grey line is the prediction for a model where the cross-interference terms are set to zero. See text for details.
	} 
\end{figure}

We compare this estimate with a numerical simulation, adding up the contributions of a chain, consisting of $N=1000$ of sodium atoms with a period $a=\lambda$, and determine the photon count signal \eqref{PCS_general} by solving the MCD equations on the laser detuning grid, $\delta_L \in \{-5 \Gamma, \delta\omega + 4 \Gamma\}$, here defined with respect to the eigenfrequency of the $R$-transition of the bare atom. To assess the influence of the motion on the signal, we focus on the CLS. At fixed $\Delta$, the spatial distribution \eqref{Gauss} is simulated by solving the system of equations \eqref{eq:MCD_chain} numerically for $N_c$ spatial configurations, according to the statistics of \eqref{Gauss}, and then averaging $S(\delta_L)$ over all configurations.  The number of spatial configurations $N_c$ is fixed in a such way that the mean squared deviation of the CLS is less than $0.001\,\Gamma$: ranging from $N_c=50$ for $\Delta=0.01\lambda$ to $N_c=500$ for $\Delta=0.1\lambda$.  The signals are fitted with the function \eqref{fano_fit} in order to extract the values of the lines' centers.  The resuls are shown in Fig.\ \ref{CLS_Na_Delta}.

In the limit of vanishing fluctuation amplitudes, $\frac{\Delta}{a} \to 0$, corresponding to the limit of the "pinned" atoms, the line shifts are the ones shown in Fig. \ref{Na_parameters_VS_period_N=1000}b at $a=\lambda$. As the width $\Delta$ of the spatial distribution increases, the absolute value of the line shift decreases. The results of the numerical simulations agree reasonably well with the predictions from the line shift rescaled by $\big(1-\frac{1}{2}(k\Delta)^2 \big)$ for widths $\Delta<0.06\lambda$, while for $\Delta \gtrsim 0.07\lambda$ there is a growing mismatch. This behavior is expected, given that Green's function \eqref{Noisy_GF}  is derived under the condition of small displacements. The convergence of the line shifts towards the result obtained without cross-interference clearly shows that motion tends to cancel the cross-interference effect. This trend highlights the increased sensitivity of cross-interference to motional noise, which must be considered in the design and performance of potential experiments.

\section{Conclusions}
\label{Sec:Conclusions}

We have analysed the impact of vacuum-induced interference on the photon-count signal from a chain of emitters, composed by two parallel dipolar transitions, which are typically discarded when describing the superradiant scattering of optical dense media. In this paper we have considered the limit in which the dipolar transitions can be treated as coherent dipoles, and showed that the interplay of Bragg scattering and multilevel interference gives rise to visible effects, which lead to a qualitative modification of the excitation spectrum. This modification becomes enhanced the smaller be the energy gap between the two dipolar transitions and/or at subwavelength distances, and leads to qualitatively different shifts and linewidths of the resonances with respect to the predictions obtained when discarding vacuum-induced cross-interference. 

In this work we took the parameters of the D2 line of alkaly-metal atoms and demonstrated that cross-interference shall be accounted for when describing scattering in optical dense media of multilevel emitters. We note that an accurate description of the spectroscopic signal shall include the full level structure of the D2 line, where the number of interfering channels is significantly larger than the reduced structure we considered. Moreover, the effects are expected to become more significant at saturation, where cross-damping terms become important.

\acknowledgements
The authors wish to thank Tom Schmit for helpful comments. This work was partly performed in the framework of the European Training
Network ColOpt, which was funded by the European Union (EU) Horizon 2020 programme under the Marie Sklodowska-Curie action, Grant Agreement 721465. GM acknowledges support from the German research Foundation (DFG RTG-3082: Engineering covalent bond in molecules and materials—Ec=m2; project 534930008).

\begin{appendix}

\section{Coarse-grained master equation and model of coherent dipoles} \label{app_mcd}

In this section we start from the coarse-grained master equation for multilevel emitters of Ref.\ \cite{konovalov2020master} and systematically derive the model of coherent dipoles for the three-level emitter by performing a perturbative expansion in the saturation parameter. 

\subsection{Coarse-grained master equation}

We report here the coarse grained master equation of Ref.\ \cite{konovalov2020master}. Let $\rho$ be the density operator for the internal states of the emitters, then its dynamics is governed by the Born-Markov master equation
\begin{equation}
\label{eq:ME}
\partial_t\hat \rho=\frac{1}{{\rm i}\hbar}[\hat H_A+\hat H_S,\hat \rho(t)]+\mathcal{L}_{D}\hat \rho(t)\,,
\end{equation}
where $\hat H_A$ is the Hamiltonian of the isolated emitters, including classical drives such as a laser, and in the absence of the coupling with the quantum electromagnetic field, while Hamiltonian $\hat H_S$ and superoperator (dissipator) $\mathcal L_D$ contain both the single-atom as well as the interatomic interference terms between parallel dipoles induced by the coupling with the quantum electromagnetic field. 

{\it Hamilton operator.} The Hamiltonian term due to the interaction with the EMF is given by the expression
$$\hat H_S=\sum_\alpha\langle \hat V_\alpha\rangle_R+\frac{1}{2}\sum_{\alpha,\beta}\hat H_{\alpha\beta}^S\,,$$
where $\langle \hat V_\alpha\rangle_R$ is the expectation value of the interactions between field and emitters and here vanishes, since we assume that the EMF is in the thermal state. The Hamilton operator $\hat H_{\alpha\beta}^S$ contains the frequency shifts and couplings due to the multilevel interference:
\begin{eqnarray} 
 \label{eq:Lamb+-}
\hat{H}_{\alpha\beta}^S = -\hbar  \sum_{i,j} &&\left[\left(\Delta_{ij}^{\alpha\beta-}+\Delta_{ij}^{\alpha\beta(T)}\right) \hat{\zeta}_{i}^{\alpha\dagger}\hat{\zeta}_{j}^{\beta}   \right.\\
&&+\left.\left(\Delta_{ij}^{\alpha\beta+}-\Delta_{ij}^{\alpha\beta(T)}\right)^*   \hat{\zeta}_{i}^{\alpha}\hat{\zeta}_{j}^{\beta\dagger}\right]+{\rm H.c.} \nonumber\,,
\end{eqnarray}
where $\hat{\zeta}_{j}^{\beta\dagger}\equiv |j_1\rangle_\beta\langle j_2|$ is the raising operator for the atom at position $\beta$ exciting the dipolar transition $j$ with state at lower energy $j_2$ and state at higher energy $j_1$ and dipole moment $\vec D_j^\beta$. The frequencies $\Delta_{ij}^{\alpha\beta(T)}=\Delta_{ij}^{\alpha\beta-}(T)-\Delta_{ij}^{\alpha\beta+}(T)$ and the individual coefficients read (below in Gauss units):
\begin{align} 
\Delta_{ij}^{\alpha\beta\pm} =&\Theta_{ij}^{(\Delta t)} \frac{\vec{D}_{i}^{\alpha*}\cdot\vec{D}_{j}^{\,\beta}}{(2\pi)^2 \hbar c^3} \mathcal{P} \int_{0}^{\omega_{\rm cut}} \frac{d\omega \,\omega^3}{\omega \pm \omega_{ij}} F_{\alpha\beta}^{ij}(\vec{R}_{\alpha\beta}), \nonumber   \\
\label{eq:coef:1}\\
\Delta_{ij}^{\alpha\beta\pm}(T)=&\Theta_{ij}^{(\Delta t)} \frac{\vec{D}_{i}^{\alpha*}\cdot\vec{D}_{j}^{\,\beta}}{(2\pi)^2 \hbar c^3}   \nonumber
\\
& \mathcal{P} \int_{0}^{\omega_{\rm cut}} \frac{d\omega \,\omega^3n(\omega,T)}{\omega \pm \omega_{ij}} F_{\alpha\beta}^{ij}(\vec{R}_{\alpha\beta})\,.\nonumber   \\    
\label{eq:coef:2} \end{align}
Here, $\mathcal{P}$ denotes the Cauchy principal value and $\omega_{\rm cut}$ is the cutoff frequency. The frequency 
$$\omega_{ij}=\frac{\bar\omega_i+\bar\omega_j}{2},$$
is the average between the two transition frequencies, and the coefficient $F_{\alpha\beta}^{ij}(\vec{R}_{\alpha\beta})$ depends also on the distance $\vec{R}_{\alpha\beta}=\vec{R}_{\alpha}-\vec{R}_{\beta}$ between the atoms and on the wave number $k=\omega/c$. It takes the form
\begin{eqnarray} \label{eq:F_diffraction_function}
&&F_{\alpha\beta}^{ij}(\vec{R}_{\alpha\beta}) = 
4\pi \left(  j_0(kR_{\alpha\beta})  \left[1- \frac{(\vec{D}_i^{\alpha} \cdot\vec{R}_{\alpha\beta})^*(\vec{D}_j^{\,\beta} \cdot\vec{R}_{\alpha\beta})}{D_i^{\alpha}D_j^{\beta}R_{\alpha\beta}^2}  \right] \right.
\nonumber\\
&&- \left. \frac{j_1(kR_{\alpha\beta})}{kR_{\alpha\beta}} 
\left[1- \frac{3(\vec{D}_i^{\alpha} \cdot\vec{R}_{\alpha\beta})^*(\vec{D}_j^{\,\beta} \cdot\vec{R}_{\alpha\beta})}{D_i^{\alpha}D_j^{\beta}R_{\alpha\beta}^2}  \right] \right)\,,
\end{eqnarray} 
where we used the notation $D_i^\alpha=|\vec{D}_i^{\alpha}|$ and $R_{\alpha\beta}=|R_{\alpha\beta}|$. Here, $j_0(x)$ and $j_1(x)$ are spherical Bessel functions of the first type. The dependence on the vector joining the atoms breaks the spherical symmetry and is at the origin of the anisotropic light emission of superradiance. 

{\it Dissipator.} The Lindblad term $\mathcal{L}_{D}$ describes the incoherent processes. It can be decomposed into the sum
\begin{equation}
\label{eq:LD}
\mathcal{L}^{D}\hat \rho(t)=\sum_{\alpha,\beta}\mathcal{L}_D^{\alpha\beta}\hat \rho(t)\,,
\end{equation}
where the terms with $\alpha=\beta$ describe the dissipation of $N$ non-interacting atoms, while the terms with $\alpha\neq\beta$ originate from multiple scattering of resonant photons and vanish when the distance between the atoms exceeds several wavelengths. The individual terms read
\begin{eqnarray}  \label{eq:dissipator}
\mathcal{L}_D^{\alpha\beta}\hat \rho(t)&=&  
 \sum_{i,j} (1+n(\omega_{ij},T))\\
& &\times\left(\frac{{\Gamma_{\alpha\beta}^{ij}}}{2}\left[\hat{{\zeta}}_{j}^{\beta} \hat{{\rho}}(t), \hat{{\zeta}}_{i}^{\alpha\dagger}\right] + \frac{{\Gamma_{\alpha\beta}^{ij}}}{2}\left[\hat{{\zeta}}_{j}^{\beta}, \hat{{\rho}}(t) \hat{{\zeta}}_{i}^{\alpha\dagger}\right] \right) \nonumber\\
&+& \sum_{i,j} n(\omega_{ij},T)\nonumber\\
& &\times\left(\frac{{\Gamma_{\alpha\beta}^{ij*}}}{2}\left[\hat{{\zeta}}_{j}^{\beta\dagger} \hat{{\rho}}(t), \hat{{\zeta}}_{i}^{\alpha}\right] + \frac{{\Gamma_{\alpha\beta}^{ij*}}}{2}\left[\hat{{\zeta}}_{j}^{\beta\dagger}, \hat{{\rho}}(t) \hat{{\zeta}}_{i}^{\alpha}\right] \right)\,, \nonumber
\end{eqnarray}
where $n(\omega,T)=1/({\rm e}^{\beta\hbar\omega}-1)$ is the mean photon number at temperature $T=1/k_B\beta$. The damping coefficients take the form
\begin{equation} \label{eq:Gamma_CG}
	\Gamma_{\alpha\beta}^{ij}=\Theta_{ij}^{(\Delta t)}\frac{\vec{D}_{i}^{\alpha*}\cdot\vec{D}_{j}^{\,\beta} }{{2\pi}\hbar c^{3}} \omega_{ij}^{3} F_{\alpha\beta}^{ij}(k_{ij})\,,  
\end{equation}
with $ k_{ij}=\frac{\omega_{ij}}{c}$. We note that for $i\neq j$ the damping coefficients are different from zero if the scalar product $\vec{D}_{i}^{\alpha*}\cdot\vec{D}_{j}^{\,\beta} \neq 0$. Finally,
\begin{equation}
\label{Eq:Theta:ij}
\Theta_{ij}^{(\Delta t)}=\frac{\sin((\omega_i+\omega_j)\Delta t/2)}{(\omega_i+\omega_j)\Delta t/2}\,,
\end{equation}
and results from the procedure of coarse graining. This term selects transitions which are resonant within the resolution set by the coarse-graining time $\Delta t$. For optical transitions, this factor selects a pair of frequencies $\omega_i$ and $\omega_j$ with opposite signs. Correspondingly, it selects terms in Hamiltonian and dissipator where the pairs of operators $\hat{\zeta}_i^{\alpha \dagger}\hat{\zeta}_j^\beta$ describe an excitation and a de-excitation along two (quasi-)resonant transitions. See \cite{buchheit2016master,konovalov2020master}.

\subsection{Derivation of the model for coupled coherent dipoles} \label{app_mcd:1}

We now consider a chain composed by $N$ identical emitters uniformly driven by the laser. We now assume that the internal levels of the emitter consist of the ground state $|g\rangle$ and excited states $|R\rangle$ and $|B\rangle$ with parallel dipolar moments $\vec{d}^R\equiv \vec{D}_{i}$ with $i=(g,R)$ and $\vec{d}^B\equiv \vec{D}_{j}$ with $j=(g,B)$. Hamiltonian $\hat H_A=\sum_\alpha \hat H_A^{\alpha}$ is the sum of the Hamiltonian for each emitter:
\begin{equation}
    \hat H_A^{\alpha}=-\sum_{R,B}\hbar \Delta^i|i\rangle_\alpha\langle i|+\hbar\left(\Omega_\alpha^i |i\rangle \langle g|+{\rm H.c.}\right)\,,
\end{equation}
and is reported in the reference frame rotating at the laser frequency $\omega_L$. Here, $\Omega_\alpha^i=\Omega^i {\rm e}^{{\rm i}\vec{k}_L\cdot\vec{R}_\alpha}$. In what follows we assume optical transitions and discard thermal effects, because the mean photon number at optical frequencies and room temperature is negligible. Therefore, we set $n(\omega_{ij},T)=0$ and consequently discard the contribution of thermal shifts $\Delta_{ij}^{\alpha\beta\pm}(T)$ to the Hamiltonian $\hat H_S$ as well as the thermal contributions to the dissipator.  

The weak saturation regime is defined by the small parameter $\mathcal{S}=N\Omega/|\Delta^i-{\rm i}\Gamma/2|\ll 1$. The state $|G\rangle=\otimes_{\alpha=1}^N|g\rangle_{\alpha} $ is the ground state of the chain and has occupation $\rho_{GG}\equiv\langle G|\hat\rho|G \rangle \sim 1$. The optical coherences between $|G\rangle$ and the excited state $|e_{\alpha}^i \rangle \equiv |g_1,\,\ldots,\,e_{\alpha}^{i},\,\ldots,\,g_N \rangle$ are $\langle e_{\alpha}^i|\hat\rho|G\rangle \sim \mathcal{S}$, while all other terms are of higher order and are here neglected. Following this procedure, the matrix elements $b^i_\alpha= \langle e_{\alpha}^i|\hat\rho|G \rangle$ (optical coherences) of master equation \eqref{eq:ME} take the form
\begin{subequations} \label{eq:cohGeneral}
	\begin{align}   
		  \dot b^i_\alpha =& -\mathrm{i}  \left(\omega^{i}
		- \omega_L-\mathrm{i} \frac{\Gamma_{\alpha}^{i}}{2}\right) b^i_\alpha -\mathrm{i} \Omega_{\alpha}^{i}\rho_{G,G} 
		\nonumber \\
		&- \sum_{\beta, \alpha=1}^{N}  \sum_{j}^{} G_{\alpha\beta}^{ij}  b_\beta^j\label{MCD}
	\end{align}
\end{subequations}
and then replace $\rho_{GG}=1$ at first order in $\mathcal S$, and refer the reader to Refs.\ \cite{pellegrino2014observation, jennewein2016coherent} for analogous treatments where the ground state is not unique. The definitions of the individual terms are reported in Sec.\ \ref{section_II}, however note that the sum over the emitters here contains the cross-damping and interference terms of the individual emitter as it contains the term $\alpha=\beta$. The term $\mathcal{F}^{ij}(\Delta t)$ approximates the function $\Theta_{ij}^{(\Delta t)}$ using the convolution with a Gaussian \cite{buchheit2016master,konovalov2020master}:
\begin{equation}
\mathcal{F}^{ij}(\Delta t) = \frac{2}{\sqrt{\pi}\tau}\int_{0}^{\infty} \frac{\sin(\frac{\tau}{2}( \omega^i-\omega^j ))}{\frac{\tau}{2}( \omega^i-\omega^j )} e^{-\frac{\tau^2}{\Delta t^2}} d\tau\,.
\end{equation}
Here $\Delta t$ is a coarse-graining  time, a time scale in the range $\tau_R \ll \Delta t \ll \tau_A$, between $\tau_R$ and the characteristic timescale $\tau_A$ of the system's internal dynamics. For degenerate dipoles this coefficient is unity. For the case $|\omega^i-\omega^j|^{-1} \gg \tau_R$, which is considered in the present work, then $\mathcal{F}^{ij}(\Delta t) \approx 1$.

\section{Level structure of line D2 of alkali atoms $^{7}$Li and $^{23}$Na} \label{app:atomic_levels}

\begin{figure}[!ht]
	\centering \includegraphics[width=1\linewidth]{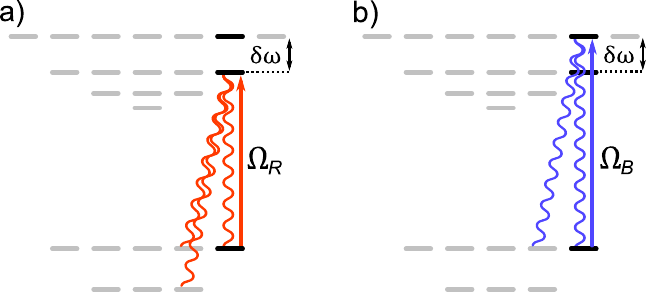}
	\caption{\label{Na23_D2} 
    Level structure for the D2 line of  $^{23}$Na atom. In the atom, initially prepared in the polarized ground state $|g^{F=2,\,M_F=2} \rangle$ and illuminated by $\pi$-polarized laser, two transitions can occur: lower frequency a) $|3S_{\frac{1}{2}}, F=2, M_F=2 \rangle \to |3P_{\frac{3}{2}}, F'=2, M_{F'}=2 \rangle$ and higher frequency b) $|3S_{\frac{1}{2}}, F=2, M_F=2 \rangle \to |3P_{\frac{3}{2}}, F'=3, M_{F'}=2 \rangle$, referred to as $R$ and $B$ in the main text. The excitations undergo the subsequent decay processes denoted with wiggling lines. In our study we ignore the decay channels to states with $M_F<2$.
	} 
\end{figure}

\begin{figure}[!ht]
	\centering \includegraphics[width=1\linewidth]{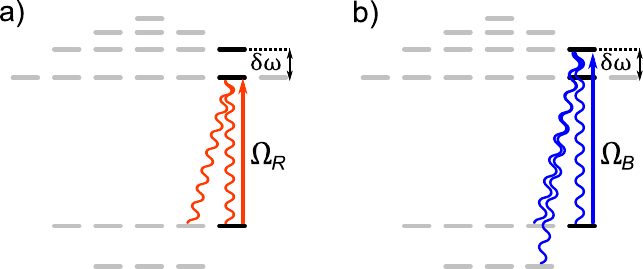}
	\caption{\label{Li7_D2} Same as Fig.\ \ref{Na23_D2} but for the level structure for the D2 line of  $^{7}$Li atom.  
	} 
\end{figure}

In this section, we provide details and parameter values for the D2 transition line, which are relevant to the calculations presented in the main text. Both isotopes, $^{7}$Li and $^{23}$Na, have a single optical electron in the outer shell and a nuclear spin $I=\frac{3}{2}$. The relevant level structures are shown in Figs. \ref{Na23_D2} and \ref{Li7_D2} for $^{23}$Na and $^{7}$Li atoms, respectively.    
For the $^{23}$Na atom the relevant dipolar transitions are: 
\begin{align}
R: \,|3S_{\frac{1}{2}}^{F=2,\,M_F=2}\rangle & \to |3P_{\frac{3}{2}}^{F'=2,\,M_{F'}=2}\rangle \nonumber
\\
B:\, |3S_{\frac{1}{2}}^{F=2,\,M_F=2}\rangle & \to |3P_{\frac{3}{2}}^{F'=3,\,M_{F'}=2}\rangle. \nonumber
\end{align}
The energy gap between the non-degenerate dipoles is taken $\frac{\delta \omega}{2\pi}= 58.326$ MHz, and the decay rate is $\frac{\Gamma}{2\pi}=9.795$ MHz \cite{steck2003sodium}.

For the atom $^{7}$Li the relevant dipolar transitions are: 
\begin{align}
R:\,|2S_{\frac{1}{2}}^{F=2,\,M_F=2}\rangle& \to |2P_{\frac{3}{2}}^{F'=3,\,M_{F'}=2}\rangle \nonumber
\\
B:\,|2S_{\frac{1}{2}}^{F=2,\,M_F=2}\rangle& \to |2P_{\frac{3}{2}}^{F'=2,\,M_{F'}=2}\rangle. \nonumber
\end{align}
The energy gap between the non-degenerate dipoles is taken $\frac{\delta \omega}{2\pi}= 9.2$ MHz, and the decay rate is $\frac{\Gamma}{2\pi}=6$ MHz \cite{zelener2015magneto}.

\end{appendix}

\bibliography{refs}

\end{document}